\documentclass[journal]{IEEEtran}
\usepackage[utf8]{inputenc}

\usepackage{xcolor}
\usepackage{color,soul}
\usepackage{amsmath}
\usepackage[nospread]{flushend}
\usepackage{url}
\usepackage{graphicx}
\usepackage{caption}
\definecolor{aqua}{rgb}{0.0, 1.0, 1.0}
\definecolor{babyblue}{rgb}{0.54, 0.81, 0.94}
\usepackage{array}

\usepackage{cite}

\ifCLASSINFOpdf
\else
\fi
\usepackage{fancyhdr}

\begin{document}
	%
	\title{Double-Sided Beamforming in OWC Systems Using Omni-Digital Reconfigurable Intelligent Surfaces}
	%
	%
	%
	
\author{Alain R. Ndjiongue, Telex M. N. Ngatched, Octavia A. Dobre, and Harald Haas 
\thanks{A. R. Ndjiongue, T. M. N. Ngatched, and O. A. Dobre are with the Faculty of Engineering and Applied Science, Memorial University, Canada. \\ Harald Haas is with the LiFi Research and Development Center, Dpt. Electronic and Electrical Engineering, the University of Strathclyde, Glasgow, U.K.}}
	\maketitle
	
	\begin{abstract}
In this paper, we introduce a variant of reconfigurable intelligent surfaces (RISs) called omni-digital-RISs (DRISs), which allow multiple physical processes, with application to optical wireless communications
systems. The proposed omni-DRIS contains both reflectors and refractive elements, as well as elements that perform both simultaneously. We describe and explain the concept of omni-DRIS, suggest and analyze an omni-DRIS coding structure, discuss metamaterials to be used, and provide a design example. Furthermore, we demonstrate that the achievable rate of an omni-DRIS system depends on the number of omni-DRIS elements, bits per phase shift, and the number of unused elements. In addition, we show that the achievable rate upper bound is related to the number of omni-DRIS elements, and conclude by discussing future research directions.
\end{abstract}
	
	\begin{IEEEkeywords}
RIS, digital-RIS (DRIS), omni-DRIS, optical wireless communications (OWC), double-side beam management, simultaneous transmission and reflection, double-sided beamforming.
	\end{IEEEkeywords}

	\IEEEpeerreviewmaketitle
	
\section{Introduction} \label{intro}
The advantages of reconfigurable intelligent surfaces (RISs) extend beyond solving the skip zone problem in wireless communication systems. Additionally, they enable wireless channels to be reconfigured and are efficient in emerging beam management. RISs are used in radio-frequency and optical wireless communication (OWC) systems (\hspace{-0.02cm}\cite{9475155,9475154, 9443170, 9614037, 9474926}, and references therein). In both reflection and refraction, the RISs can perform beam steering, splitting, scattering, light polarization, and photon absorption \cite{9475154}. To solve skip zones in wireless systems, an RIS is placed over the wireless channel and becomes a part of it. By adjusting its physical properties, the channel can be reconfigured. As a result, the RIS can resolve dead zones, create reconfigurable channels, and manage emerging beams \cite{9475160}. In this position, the RIS module is mainly used to reflect the incident signals. Therefore, both incident and reflected signals remain in the same half-space with respect to the RIS. New development have proven that the RIS module can be mounted inside the transmitter or receiver \cite{9354893, 9690473}. In this configuration, the RIS allows the signal to pass through it and acts as a refracting system that can steer the impinging signal and control the emerging beams. Though the emerging signal lies in a different half-space, the RISs elements deal with a unique physical process, and are not digitized.

As a result of the more recent development in the RIS technology, digital-RISs (DRISs), which allow digital signal processing to be applied to the physical material used in the reconfigurable elements, have been conceived \cite{ndjiongue2021digital}. The entire system is governed by a control module, which generates the code sequence for each element. According to the feedback from users, the control module of the DRIS transfers a specific code to the targeted element. During the execution of the code, physical properties of the DRIS elements are reconfigured to fulfill the new wave propagation required by the users. 

Simultaneous transmission and reflection (STAR) for 360$^o $ coverage by RISs has also been demonstrated recently \cite{9690478}. STAR-RISs allow a double physical processes at the reconfigurable elements. Some of the elements act as reflectors, while others guide the signal through the structure. The remaining elements may reflect and refract light simultaneously or remain unused. Therefore, to connect users located on both sides, the RIS module must be equipped with both reflectors and elements that can refract signals, as well as elements that can perform the two processes simultaneously \cite{9690478}. 

In this paper, we introduce a RIS that combines the characteristics of STAR-RISs and DRISs with application to OWC systems. The proposed RIS, referred to as omni-DRIS can achieve both reflection and refraction simultaneously, while enabling  digital signal processing in its elements. It is thus able to spatially direct light toward users located on both sides of the module. We discuss the omni-DRIS, propose and investigate a coding protocol, review metamaterials that can be used with an emphasis on liquid crystals (LCs), and propose double-sided beam management for omni-DRIS-assisted OWC systems. To conclude, we consider a case study of an omni-DRIS-assisted indoor visible light communication (VLC) design example, and through the achievable rate, we analyze the system's performance with regard to the number of active, inactive, and total elements of the omni-DRIS.

\section{Ordinary RIS vs. Omni-DRIS} \label{loss}
When deployed, the RIS has always been mounted on a building wall, pasted to a flying vehicle, or incorporated into the device's transmitter or receiver \cite{9614037, 9474926, 9354893}. In such configurations, the RIS is composed of elements that perform only one physical process at a time. Specifically, they serve as reflectors or refractive devices. These configurations prevent the user located at the back of the RIS module or inside the building from being linked to the network at the same time as the users on the other side of the RIS \cite{9690478}. Thus, the RIS creates a blind spot that may be resolved in two ways: (\textit{i}) by utilizing a second RIS to cover the back of the first, or (\textit{ii}) by using an omni-RIS module, which incorporates elements that can serve as both reflectors and refractive devices simultaneously \cite{9690478,9491943}. The latter solution has the additional advantage of being economical. The proposed omni-DRIS is comprised of elements that can be in one of four possible states: reflection, refraction, both reflection and refraction, and unused. Reflectors reflect the light impinging on the omni-DRIS surface towards the user located on the same side of the incident light. The element has a reflection coefficient close to unity and a phase shift between zero and pi. On the other hand, an omni-DRIS element that acts as a refractive device allows the light ray to pass through it and is characterized by a high transmission coefficient close to one and a phase shift ranging between pi and two pi. An omni-DRIS element which simultaneously reflects and refracts the incoming light ray combines the characteristics of a reflector and refractive device. Note that in this case, the sum of the refraction and reflection coefficients equals unity. Lastly, the omni-DRIS element can remain quite (or turned off).
\begin{figure}
	\centering
	\includegraphics[width=0.48\textwidth]{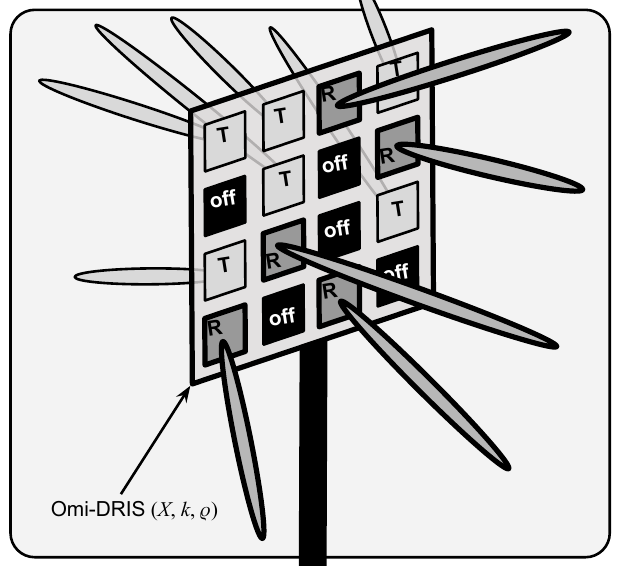}
	\caption{An illustration of the omni-DRIS operation. A number of elements are reflective while some are refractive, and a set of elements is off and will be used as soon as needed.}
	\label{fig:odris}
		\vskip -1\baselineskip plus -1fil
\end{figure}

\begin{figure}
	\centering
	\includegraphics[width=0.48\textwidth]{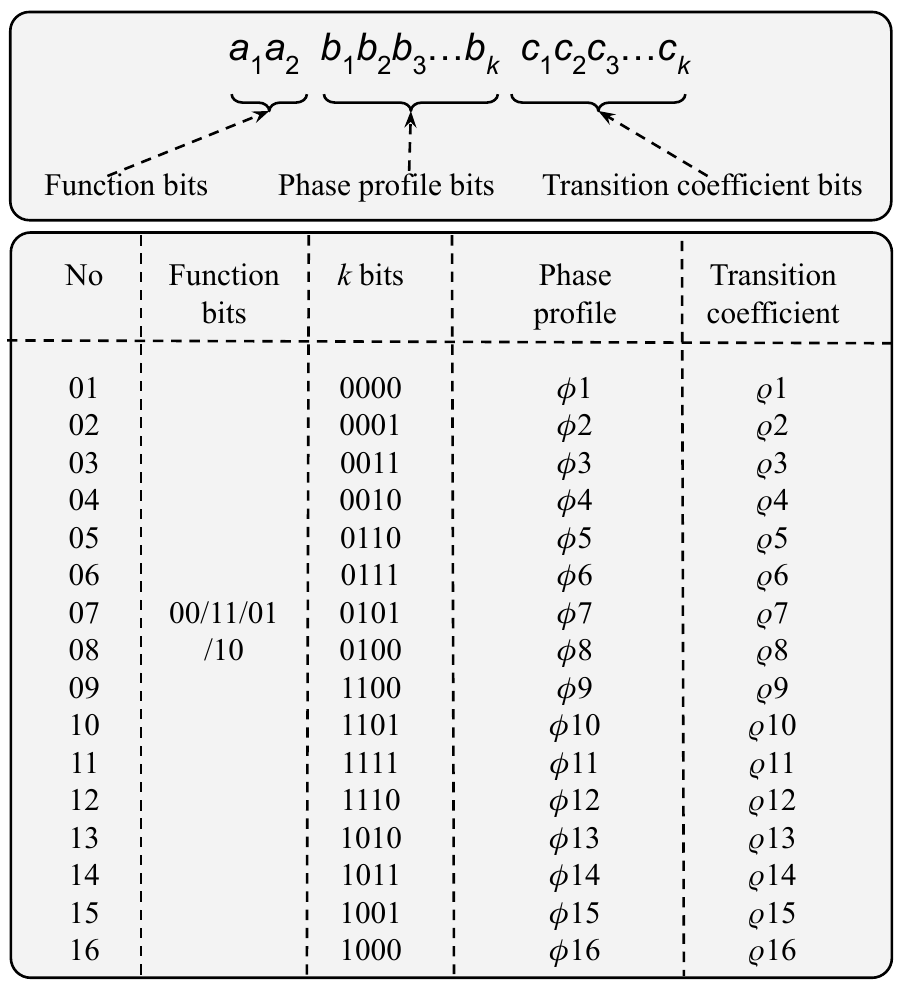}
	\caption{Structure of the omni-DRIS codes (top). Omni-DRIS mapping example: $ k $ = 4, 16 omni-RIS element types (bottom).}
	\label{fig:coding}
	\vskip -1\baselineskip plus -1fil
\end{figure}

Figure~\ref{fig:odris} illustrates an omni-DRIS, showing a double-sided beam generation. It reflects and refracts light simultaneously within the system in which it is inserted. Elements marked with an "R" are total reflectors, while those marked with a "T" totally refract the incident light. Elements marked with "OFF" are not in use. To avoid having an overcrowded figure, this illustration does not show reflective-refractive elements. For similar reasons, the incident light beams are not shown. To implement the omni-DRIS, an algorithm must be developed that controls each element individually. The generated codes are transferred to the module as sequences of bits. This aspect is discussed in the next section.
\begin{figure*}
	\centering
	\includegraphics[width=0.85\textwidth]{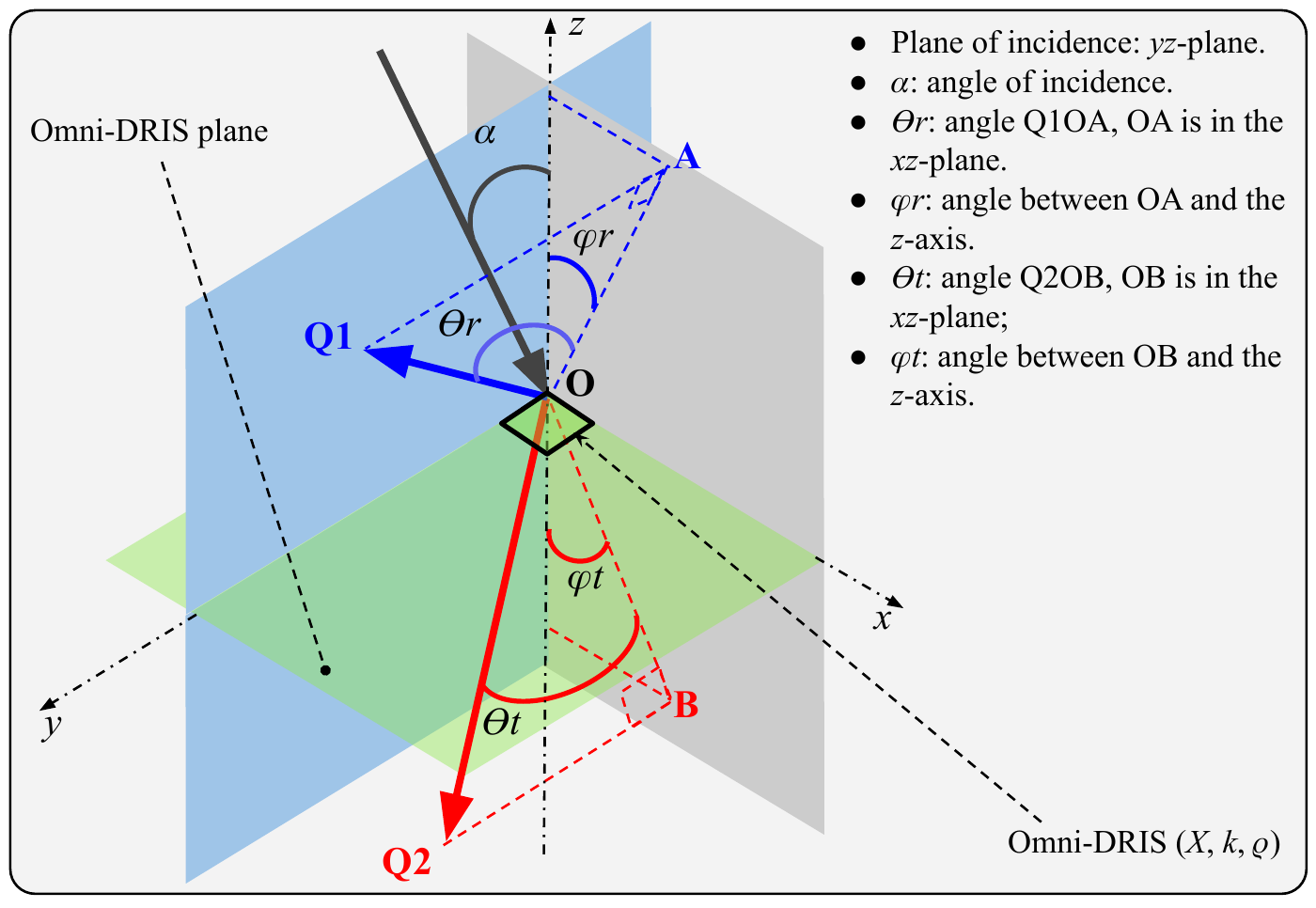}
	\caption{Illustration of light reflection and refraction through the omni-DRIS in three dimensions.}
	\label{fig:3D_Syst}
	\vskip -1\baselineskip plus -1fil
\end{figure*}
\section{Coding Omni-DRIS and Emerging Light Construction}\label{Section:three}
Similar to the DRIS proposed in \cite{ndjiongue2021digital}, the omni-DRIS is composed of $ 2^k $ elements, where $ k $ refers to the number of bits necessary to define the phase shifts and transition coefficients. Therefore, the omni-DRIS is composed of $ 2^k $ phase shifts and transition coefficients. To control the omni-DRIS, a series of bit sequences, having the same length, is mapped to the phase shifts and transition coefficients of its elements. Each code sequence is composed of three blocks. The coding structure is depicted at the top of Fig.~\ref{fig:coding}, where the first block is $ a_1a_2 $, while $ b_1b_2b_3\dots b_k $ and $ c_1c_2c_3\dots c_k $ are the second and third blocks, respectively. The first block, made of two bits, is used to define the state of the omni-DRIS element. "01" and "10" respectively induce reflection and refraction, "11" is reserved for an element with two-process capability, i.e., simultaneous reflection and refraction, while "00" is reserved for an omni-DRIS element that should remain off. The second block, made of $ k $ bits, defines the phase profiles, and the third block, also made of $ k $ bits, is reserved for the transition coefficients. As a result, each code is made of 2($ k $ + 1) bits. For example, an omni-DRIS with 16 active elements has a code sequence composed of 10 bits. Two bits to select the element's mode, four bits to define the phase shifts, and four bits to represent the transition coefficients. 

The bottom part of Fig.~\ref{fig:coding} illustrates a mapping table for a coding scheme for an omni-DRIS, which contains 16 types of active elements. The first column indicates the code number, the second column contains two bits to induce the physical process, "01", "10", "11", and "00". The third column encloses a four-bit gray coding arrangement for defining the phase profiles and transition coefficients. It is pertinent to recall that the expression "phase shift" is only used if the reflected or refracted light lies over the same plane as the incident light ray. Due to the fact that the plane of the emerging light may differ depending on the position of the user, which may be mobile, the emerging light is not always in the plane of incidence. As a result, the values in column three represent the phase profiles of the omni-DRIS elements. The phase profile is the result of combining the two phase shifts that represent the off-plane orientation of the emitted light. For instance, if the incident light lies in the $yz$-plane, the phase profile of the emerging light is given with reference to that plane. The phase profile, shown at the bottom of Fig.~\ref{fig:coding}, groups the angle between the emerging light and the $xz$-plane, as well as the angle formed by the projection of the emerging light onto the $xz$-plane and the $z$-axis.

Figure~\ref{fig:3D_Syst} depicts an illustration of such three-dimensional light orientation. The figure shows the transition of a light ray through an omni-DRIS element, highlighting both reflection and refraction for the same incident light ray. The values of phase profiles, given in Fig.~\ref{fig:coding}, comprise a combination of $\theta_r$ (angle between the emerging light ray and the $ xz $-plane) and $\varphi_r$ (angle between the projection of the emerging light ray to the $ xz $-plane and the $ z $-axis) for reflection, as well as a combination of $\theta_t$ (angle between the emerging light ray and the $ xz $-plane) and $\varphi_t$ (angle between the projection of emerging light ray to the $ xz $-plane and the $ z $-axis) for transmission. Accordingly, when $yz$ is the plane of incidence, the omni-DRIS element directs the light in a specific direction defined in relation with the $xz$ and $yz$ planes, when the omni-DRIS is in the $xy$ plane. Likewise, for the same plane of incidence and omni-DRIS plane as above, the refracted ray direction is given with respect to both the $xz$ and $yz$ planes. When the impinging light remains in the $yz$-plane, the emerging light can be described as either reflection or refraction if it lies over the omni-DRIS's plane and forms a 90-degree angle with the $ z $-axis. In this case, the phase profile of the omni-DRIS element is the angle formed by the emerged ray and one of the $x$ or $y$ axes. As previously presented, the path of light through an omni-DRIS element and the corresponding phase profile, taken together with the sequences of control codes, demonstrate that an omni-DRIS steers the light with respect to a specific three-dimensional constellation. Since the number of phase profiles equals $ 2^k $, the number of directions is determined by the value $k$ in this constellation. The combination of the above description and mapping procedure of the bit sequences to the omni-DRIS elements, along with the concept of spatial three-dimensional emerging light, leads to two-sided beam management. An example is discussed in the following section along with a description of reconfigurable metamaterials for omni-DRIS. 
\section{Omni-DRIS Design for Double-Sided Beam Management}
\subsection{Reconfigurable Metamaterials for Omni-DRIS}
Several metamaterials may be used to manufacture omni-DRIS modules for OWC systems. 

Remote control of omni-DRIS-assisted OWC systems is a desirable feature for a variety of reasons. This control can be achieved through external parameters such as an electric or magnetic field. Thus, the omni-DRIS reconfigurable metamaterials should be selected based on their tunability and reaction to external parameters applied remotely. Therefore, electro-optical metamaterials are the perfect candidates. These includes plasmonic metamaterials and two-state materials, to mention only two. 

Plasmonic metamaterials can be classified into passive and active metamaterials. The application discussed in this paper calls for active plasmonic metamaterials since they provide the required optical properties. Active metamaterials are mostly exploited in optical, thermo-optical, and electro-optical systems \cite{ElectroNature}. These materials have the properties necessary to be used in information processing, communication, and data storage. Example of plasmonic materials include glass, prism, and lenses. Metamaterials with negative and positive indices are needed to respond to emerging light's three-dimensional and spherical orientation. Two-state materials may also be used. An example of two-state metamaterial is the chalcogenide alloy such as Germanium telluride, which exhibits high resistance in the off-state and low resistance in the on-state \cite{9475154}. Phase transitions in this material can be controlled by external factors such as electrical current or light. These materials can be used in systems requiring binary metamaterials that operate in "on" and off" phases. Due to their absorption capabilities, two-dimensional nanosheets such as dichalcogenides are also highly used in optical applications \cite{2Dchalco}. On the other hand, surface plasmons exhibit positive real permittivity, which plays an influential role in enhancing the Raman spectroscopy \cite{Raman}. This advantage, in addition to its tunability, justifies its use in the manipulation of light beams. 

Despite the advantages offered by the above metamaterials, anisotropic materials may be better suited for omni-DRIS. Unlike isotropic metamaterials, which exhibit the same properties in all directions, anisotropic metamaterials exhibit different properties in different directions within the material structure \cite{1284588}. The anisotropic behavior is most frequently observed in crystals of solid or compound materials, but rarely in liquids. Mixtures of liquids and crystals provide the properties of the liquid, and the resulting product is anisotropic if the employed crystals are also anisotropic. Anisotropic materials are characterized by phenomena such as double refraction or birefringence, a phenomenon that results in the change of light speed along the axis of the crystal. The anisotropic behavior can be observed in Selenium's electrical resistivity, which exhibits high or low values according to the externally applied field polarity. Liquid crystals (LCs) are anisotropic materials with a birefringence that can be controlled externally and remotely.\footnote{It should be noted that an LC is a state of matter that has both liquid and solid crystal properties.} Due to its electrically controllable birefringence, the light's path through an LC substance may be controlled smoothly. This control is based on varying the externally applied field, which changes to a specific value that is allocated to the corresponding element, leading to a birefringence (refractive index) variation. This procedure results in the element acquiring a phase profile and transition coefficient, causing the impinging light to be refracted or reflected in a specific direction with a specific power. 

LC-based metasurfaces possess a unique advantage. The same LC's substance and dye can be used in devices doing both reflection and refraction. Specifically, the bottom glass layer of the LC-based metasurface structure can be replaced with a silicon to convert it from a refractive device to a reflector. Furthermore, utilizing transparency-configurable materials such as changing acrylic metamaterials at the bottom layer of an LC-based metasurface yields a structure that can simultaneously reflect and refract light. The LCs' birefringence (refractive index) is sensitive to the variation of the externally applied electric or magnetic field. As a consequence, it is easy to control the light's path through LCs \cite{9354893, 9690473}. At a given temperature, varying the externally applied field induces a retardation angle, which is responsible for the shift created on the impinging light. The omni-DRIS is structured in such a way that each element is controlled individually. Therefore, it is convenient to control each omni-DRIS element using a sequence of bits. In this manner, the omni-DRIS elements can be identified with a unique code sequence as explained in Section~\ref{Section:three}. Each code sequence corresponds to a distinct type of physical process (reflection vs. refraction, or both), a phase profile (azimuth angle and elevation angle, both from zero to two pi), and a transition coefficient. The elevation angle in the phase profile varies between zero and pi for reflection and pi to two pi for refraction, while the azimuth angle varies from zero to two pi in both reflection and refraction. In view of this benefit, LCs are adopted in this paper as primary materials for omni-DRIS. As such, LCs can be used in reflective and refractive elements, and in elements that simultaneously provide the two processes, as elaborated subsequently.

\begin{figure*}
	\centering
	\includegraphics[width=0.89\textwidth]{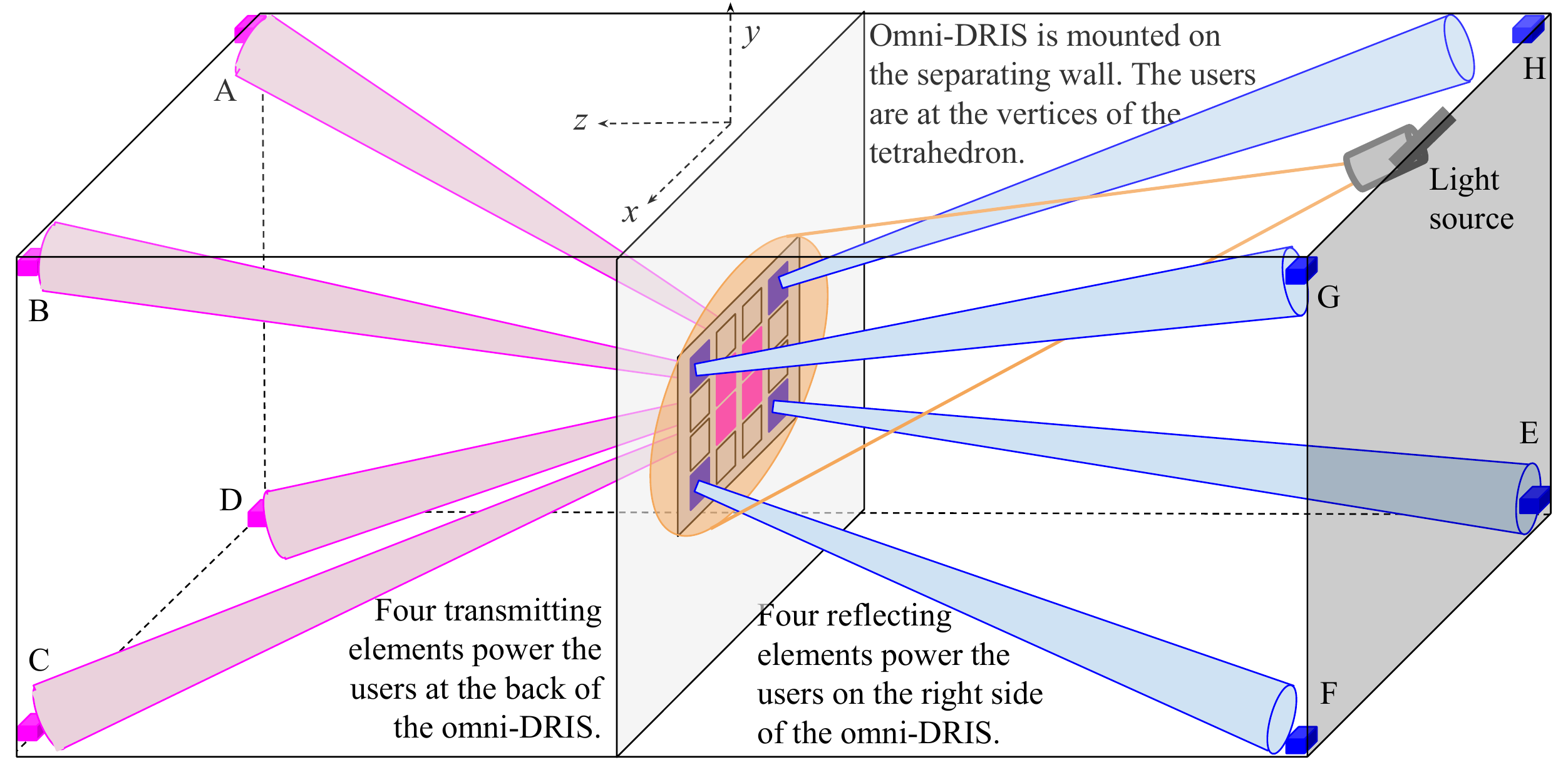}
	\caption{An example of scenario where the omni-DRIS can be used. The omni-DRIS is part of the wall separating the two parts of a room and the information light source is located in the first part. Four users are located at A, B, C, and D in Part 1 and four other users are situated at E, F, G, and H in Part 2. Each omni-DRIS element serves a different user.}
	\label{fig:3D_Example}
	\vskip -1\baselineskip plus -1fil
\end{figure*}
\noindent
\textbf{As reflective elements}: An LC sandwiched between two polarizers is an effective refractive device. To reflect the incident light, it is necessary to add an opaque layer after the bottom polarizer \cite{9354893}. To control the emerged light and spatially orient it in the desired direction, a discontinuity should be created on the upper face of the bottom opaque layer. This can be achieved by using a blazed grating system in three dimensions. Silicon is generally used as an opaque layer \cite{9690473}. Thus, smooth control can be realized on both the phase profile and the reflection coefficient by utilizing both the birefringence and the grating system. The final material is a pure reflector and cannot allow the light through the structure.  

\noindent
\textbf{As transmissive elements}: In transmittance, the material is similar to that used in reflectors, with the exception that the opaque bottom layer is replaced by a glass. To control the off-axis orientation of the transmitted light, a discontinuity must be created on either the lower or upper faces of the glass. Consequently, the emerged light can be easily controlled with the grating system and by tuning the LC's birefringence. 

\noindent
\textbf{As element performing the two processes}: An invaluable function of some omni-DRIS elements is their ability to reflect and refract light at the same time. These elements combine the features of a reflector and refractive device. However, it is vital to keep in mind that between reflection and refraction, a ratio of proportionality should be respected. If, for example, 60\% of the light impinging on the omni-DRIS surface is transmitted, then only 40\% of this light will be reflected. The concern is focused on the bottom layer of the omni-DRIS element. It should be possible to control the transparency property of the material that creates the discontinuity in the element's bottom layer. Innovative materials, such as smart and switchable glasses, can be used because they can be electronically adjusted between total transparency and opacity. The transparency of some materials, such as changing acrylic, can be controlled by heating. These materials (smart glasses and changing acrylic materials) display varying levels of transparency and opacity based on the control parameter selected. 
\subsection{Design Procedure and a Mapping Example}
The design involves defining three-dimensional constellations of the omni-DRIS and determining the corresponding code sequences. Consider an omni-DRIS composed of $ N $ elements that aim to direct an incident light to $ N $ points. This three-dimensional constellation consists of a three-dimensional geometric figure with $ N $ vertices. $ N $ is a power-of-two, 2$ ^k $, as in the modulation schemes (phase profiles and transition coefficient in the case of omni-DRIS). This geometric figure is constructed in a three-dimensional space ($ x $, $ y $, $ z $), where $yz$ is the incident plane and the users are located at its vertices. $ xy $ is the omni-DRIS plane and its elements are anisotropic LCs made of the same substance and dye. However, the tuning of each element is based on a unique control code sequence.

In the illustration provided in Fig.~\ref{fig:3D_Example}, $ k $ = 3, $ N $ = 8, the users are placed in a way to obtain a tetrahedron or cuboid, in a room divided in two parts. The omni-DRIS is part of the separating wall, and the light source carrying the message is located in the first part. The omni-DRIS contains 16 elements, with eight unused elements (no bit sequences are transferred to them and no electric field is applied). The remaining eight elements are organized such that four at the four corners of the arrangement are used as reflectors, while the four located at the center are utilized to refract the incident light. The reflective elements are LC on silicons and the upper face of each element is a grating system with a specific discontinuity. On the other hand, the elements exploited to target the users at the back of the omni-DRIS have a transparent glass at their bottom layer with a discontinued grating system. The same LC and dye are used in all elements; however, each of them is controlled individually. The tuning is coordinated locally at the omni-DRIS based on the different code sequences transferred to it. With reference to the order provided at the bottom of Fig.~\ref{fig:coding} and the element arrangement in Fig.~\ref{fig:3D_Example}, codes no. 1, 4, 13, and 16 are transferred to the outer elements, while codes no. 6, 7, 10, and 11 are transferred to the inner elements. These codes are generated according to the protocol proposed at the top of Fig.~\ref{fig:coding}. Thus different codes are transferred to the eight active omni-DRIS elements. These codes are typically designed based on the feedback from the corresponding users. Their actions on the elements are to adjust the externally applied electric field. This creates specific values of the element's temperature and refractive index, leading to an azimuth deviation and elevation in the emerging light orientation. The obtained temperature and refractive index yield a phase profile containing the corresponding elevation and azimuth angles. In this practical example, codes 0100001000, 0100101000, 0110101000, and 0110001000 respectively yield (31.22, -27.39), (-31.22, -27.39), (31.22, 27.39), and (-31.22, 27.39) for reflectors, while codes 1001111000, 1001011000, 1011011000, and 1011111000 yield (36.47, -30.33), (36.47, 30.33), (-36.47, -30.33), and (-36.47, 30.33), respectively, for the refractive elements.\footnote{Note that in the pair ($\theta$, $\varphi$), $\theta$ and $\varphi$ are the elevation and azimuth angles, respectively.} The dimensions of the beam formed by these omni-DRIS elements depends on the reflective surface and the discontinuity orientation of the elements. Thus, the narrowness of the emerged beam relates to the physical structure of the elements, most precisely the reflecting area of the element. Described above is an example of double-sided beam management in indoor VLC systems. In this OWC beamforming, the motion of the user is followed by a change in the omni-DRIS code, which results in an adjustment of the azimuth and elevation angles to track the user. In cases where the user is out of reach, beam switching prevents the user from losing connection. 

\begin{figure}
	\centering
	\includegraphics[width=0.49\textwidth]{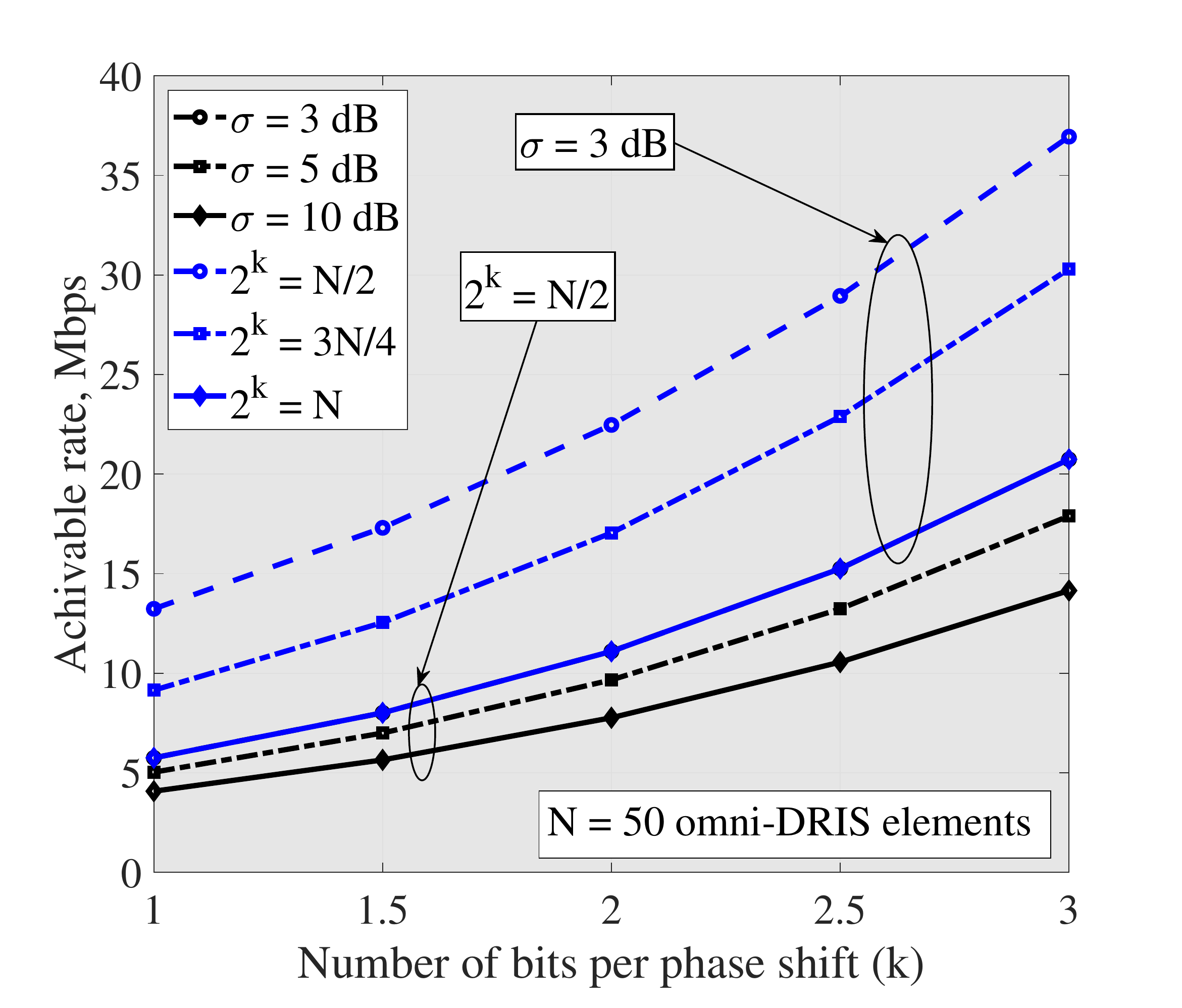}
	\caption{Achievable rate of an omni-DRIS assisted indoor VLC system in terms of bits per phase shift for a variety of noise levels.}
	\label{fig:Rate}
		\vskip -1\baselineskip plus -1fil
\end{figure}

\begin{figure}
	\centering
	\includegraphics[width=0.49\textwidth]{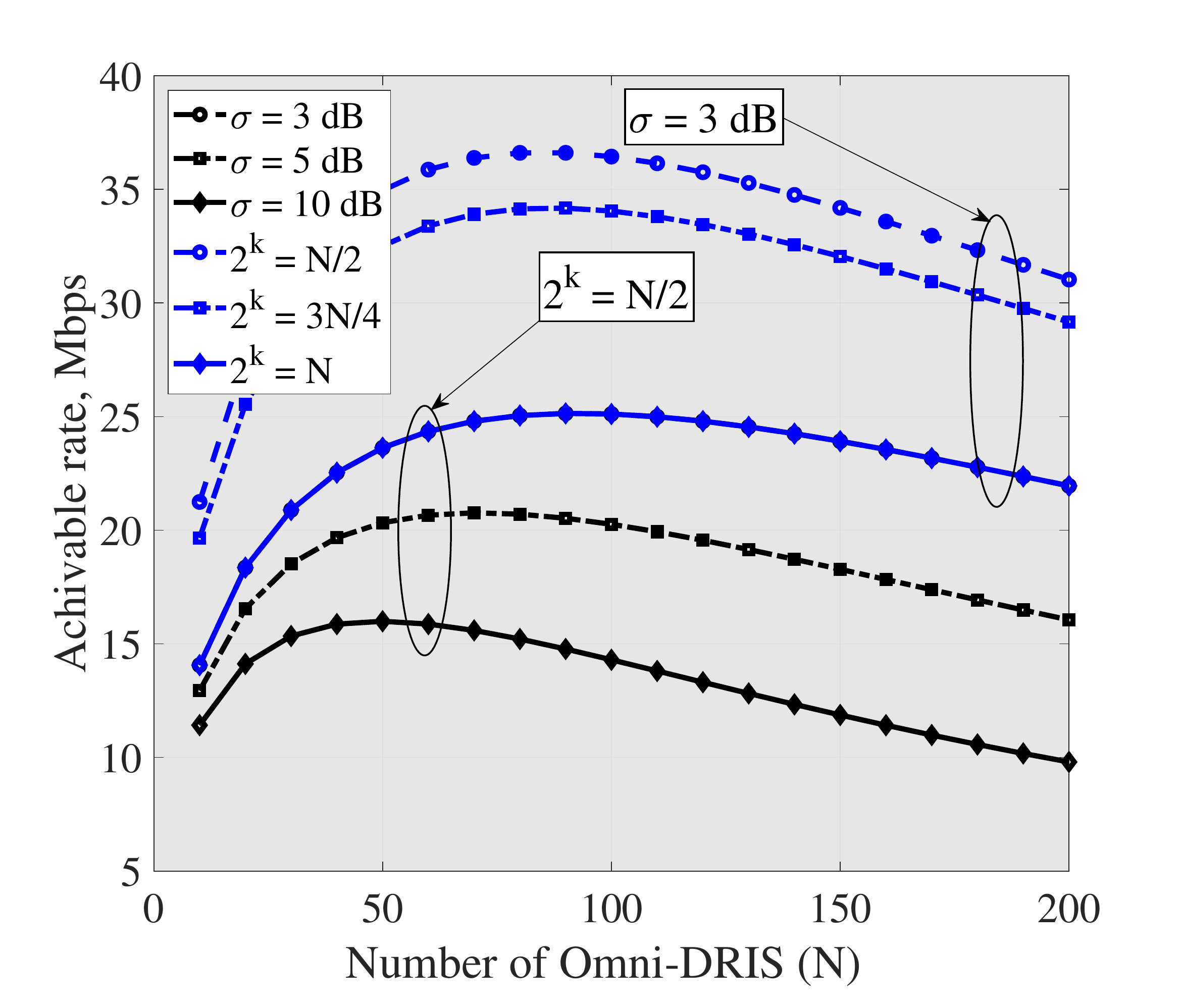}
	\caption{Achievable rate of an omni-DRIS assisted indoor VLC system in terms of the number of omni-DRIS elements for a variety of noise levels.}
	\label{fig:Rate_N}
	\vskip -1\baselineskip plus -1fil
\end{figure}

It is useful to establish threshold ratios between the number of omni-DRIS elements, the elements that are used, and those that are not used. These ratios impact the achievable rate of the system as shown in Fig.~\ref{fig:Rate} in terms of the number of bits per phase shift (transition coefficient). We demonstrate in this figure that the achievable rate is inversely proportional to the number of elements used, $ 2^k $, compared with a fixed total number of omni-DRIS elements. Note that the behavior observed in Fig.~\ref{fig:Rate} relates to the number of users. Additionally, Fig.~\ref{fig:Rate} implies that as the number of bits per phase shift (transition coefficient) increases, the system's achievable rate increases, which may suggest that the system will be able to perform at a higher rate for an infinite amount of omni-DRIS elements. However, this is not the case. It is pertinent to note that there is an upper limit, i.e., there is a number of omni-DRIS elements which results in the highest achievable rate. Over this threshold, the achievable rate is reduced as the number of bits per phase shift increases. This is illustrated in Fig.~\ref{fig:Rate_N}, where the achievable rate is given in terms of the number of omni-DRIS elements. The figure shows that there is a value for the number of omni-DRIS elements where the system has the maximum achievable rate.
\section{Future Research Directions and Conclusion}
\noindent
\textbf{Geometrical analysis of omni-DRIS systems and double-sided beamforming}: The development of an omni-DRIS modules for OWC systems may have a variety of applications as underlined in \cite{9690478} for the STAR-RIS. However, this idea is still in its infancy and therefore requires extensive and intensive investigation. It is important to identify these applications and analyze their respective channels and performances. Double-sided beamforming in omni-DRIS-assisted OWC systems, including the indoor VLC and free-space optical communication environments should be studied.

\noindent
\textbf{Omni-DRIS-assisted integrated sensing and communication (ISAC)}: The integration of sensing and communication technologies to utilize wireless resources efficiently and potentially avail of their mutual benefits is one of the major features of 5G-advanced and 6G mobile networks. The proposed omni-DRIS can find a wide range of applications in ISAC, where the sensing signal can be used to design the control sequence of the different elements.   

\noindent
\textbf{Omni-DRIS-assisted OWC system models and channel estimation}: Modeling an OWC channel assisted by an LC-based omni-DRIS has many challenges. First, when the omni-DRIS is exploited as a reflective device, the communication signal transits through the LC substance with a double run. In addition, the double refraction at the surface of the omni-DRIS will have an impact on the transmitted message. Second, the transmissive omni-DRIS element oversees the light path in a single run and a double phase shift at its input (top) and output (bottom), leading to the transmitted signal being affected. Third, both influences related to reflection and refraction will simultaneously happen to the impinging signal while facing a transmissive-reflective omni-DRIS element.

\noindent
\textbf{Omni-DRIS-assisted OWC systems implementations}: Practical deployment of an omni-DRIS-assisted OWC system requires an efficient LC materials selection. An analysis of the transition time to meet the 5G and beyond transmission speed requirement is needed. Apart from these, the spherical orientation of the emerging light in these applications requires specific materials to meet the omni-orientation of the light beams on both side of the module. Transferring the code sequences from the control module to the omni-DRIS elements have not yet been studied. The algorithm that will perform the omni-DRIS element's tuning and reconfiguration is also to be designed.  

In this paper, we investigated the use of omni-DRIS in OWC systems. Its concept was described and explained. A number of applications of this type of DRIS were also highlighted. Additionally, we examined the omni-DRIS coding and outlined a code structure. We also explored the omni-DRIS design by emphasizing the materials that could be utilized as reflectors, to transmit a light, and both phenomena simultaneously. A proposal was made to use an LC-based metasurface to realize the three types of materials by altering its bottom layer, and a design example was investigated. 
     
	\ifCLASSOPTIONcaptionsoff
	\fi
	\bibliographystyle{IEEEtran}
	\bibliography{ODRIS}

\begin{thebibliography}{10}
\providecommand{\url}[1]{#1}
\csname url@samestyle\endcsname
\providecommand{\newblock}{\relax}
\providecommand{\bibinfo}[2]{#2}
\providecommand{\BIBentrySTDinterwordspacing}{\spaceskip=0pt\relax}
\providecommand{\BIBentryALTinterwordstretchfactor}{4}
\providecommand{\BIBentryALTinterwordspacing}{\spaceskip=\fontdimen2\font plus
\BIBentryALTinterwordstretchfactor\fontdimen3\font minus
  \fontdimen4\font\relax}
\providecommand{\BIBforeignlanguage}[2]{{%
\expandafter\ifx\csname l@#1\endcsname\relax
\typeout{** WARNING: IEEEtran.bst: No hyphenation pattern has been}%
\typeout{** loaded for the language `#1'. Using the pattern for}%
\typeout{** the default language instead.}%
\else
\language=\csname l@#1\endcsname
\fi
#2}}
\providecommand{\BIBdecl}{\relax}
\BIBdecl

\bibitem{9475155}
{G. {Alexandropoulos} {\textit{et al.}}}, ``{Reconfigurable intelligent
  surfaces for rich scattering wireless communications: Recent experiments,
  challenges, and opportunities},'' \emph{IEEE Commun. Mag.}, vol.~59, no.~6,
  pp. 28--34, Jun. 2021.

\bibitem{9475154}
{C. {Molero} {\textit{et al.}}}, ``{Metamaterial-based reconfigurable
  intelligent surface: 3D meta-atoms controlled by graphene structures},''
  \emph{IEEE Commun. Mag.}, vol.~59, no.~6, pp. 42--48, Jun. 2021.

\bibitem{9443170}
{M. {Najafi} {\textit{et al.}}}, ``{Intelligent reflecting surfaces for free
  space optical communication systems},'' \emph{IEEE Trans. Commun.}, vol.~69,
  no.~9, pp. 6134--6151, Sep. 2021.

\bibitem{9614037}
{H. {Abumarshoud} {\textit{et al.}}}, ``{LiFi through reconfigurable
  intelligent surfaces: A new frontier for 6G?}'' \emph{IEEE Veh. Technol.
  Mag.}, pp. 2--11, Nov. 2021.

\bibitem{9474926}
{A. R. {Ndjiongue} {\textit{et al.}}}, ``{Toward the use of re-configurable
  intelligent surfaces in VLC systems: Beam steering},'' \emph{IEEE Wireless
  Commun. Mag.}, vol.~28, no.~3, pp. 156--162, Jun. 2021.

\bibitem{9475160}
{C. {Pan} {\textit{et al.}}}, ``{Reconfigurable intelligent surfaces for 6G
  systems: Principles, applications, and research directions},'' \emph{IEEE
  Commun. Mag.}, vol.~59, no.~6, pp. 14--20, Jun. 2021.

\bibitem{9354893}
{A. R. {Ndjiongue} {\textit{et al.}}}, ``{Re-configurable intelligent
  surface-based VLC receivers using tunable liquid-crystals: The concept},''
  \emph{IEEE/OSA J. Lightw. Technol.}, vol.~39, no.~10, pp. 3193--3200, May
  2021.

\bibitem{9690473}
------, ``{Design of a power amplifying-RIS for free-space optical
  communication systems},'' \emph{IEEE Wireless Commun. Mag.}, vol.~28, no.~6,
  pp. 152--159, Dec. 2021.

\bibitem{ndjiongue2021digital}
------, ``{Digital RIS (DRIS): The future of digital beam management in
  RIS-assisted OWC systems},'' \emph{arXiv}, Dec. 2021.

\bibitem{9690478}
{Y. {Liu} {\textit{et al.}}}, ``{STAR: Simultaneous transmission and reflection
  for 360° coverage by intelligent surfaces},'' \emph{IEEE Wireless Commun.
  Mag.}, vol.~28, no.~6, pp. 102--109, Dec. 2021.

\bibitem{9491943}
{S. {Zhang} {\textit{et al.}}}, ``{Intelligent omni-surfaces: Ubiquitous
  wireless transmission by reflective-refractive metasurfaces},'' \emph{IEEE
  Trans. Wireless Commun.}, vol.~21, no.~1, pp. 219--233, Jan. 2022.

\bibitem{ElectroNature}
{I. C. {Benea-Chelmus} {\textit{et al.}}}, ``{Electro-optic spatial light
  modulator from an engineered organic layer},'' \emph{Nat. Commun.}, vol.~12,
  no.~1, pp. 1--10, Oct. 2021.

\bibitem{2Dchalco}
{T. {Mueller} {\textit{et al.}}}, ``{Exciton physics and device application of
  two-dimensional transition metal dichalcogenide semiconductors},'' \emph{Nat.
  2D Mater Appl.}, vol.~2, no.~1, pp. 1--12, Sep. 2018.

\bibitem{Raman}
{X. {Wang} {\textit{et al.}}}, ``{Fundamental understanding and applications of
  plasmon-enhanced Raman spectroscopy},'' \emph{Nat. Rev. Phy.}, vol.~2, no.~5,
  pp. 253--271, Sep. 2020.

\bibitem{1284588}
{M. {Bullo} {\textit{et al.}}}, ``{Isotropic and anisotropic electrostatic
  field computation by means of the cell method},'' \emph{IEEE Trans. Magn.},
  vol.~40, no.~2, pp. 1013--1016, Mar. 2004.

\end{thebibliography}
\begin{IEEEbiographynophoto}{Alain R. Ndjiongue}[S'14, M'18, SM'20]
is a senior researcher at Memorial University of Newfoundland, Canada. 
\end{IEEEbiographynophoto}
\begin{IEEEbiographynophoto}{Telex. M. N. Ngatched} [M’05, SM’17]
is an associate professor at Memorial University, Canada.
\end{IEEEbiographynophoto}
\begin{IEEEbiographynophoto}{Octavia A. Dobre} [M’05, SM’07, F’20]
is a professor and research chair at Memorial University, Canada. 
\end{IEEEbiographynophoto}
\begin{IEEEbiographynophoto}{Harald Haas} [S’98, AM’00, M’03, SM’16, F’17]
holds the Chair of Mobile Communications at the University of Strathclyde.
\end{IEEEbiographynophoto}	
\end{document}